# Heterogeneity in Tribologically Transformed Structure (TTS) of Ti-6Al-4V under fretting

**Vivien Lefranc[a, b], Soha Baydoun[b], Camille Gandiolle[a], Eva Héripré[a], Maxime Vallet[a, c], Siegfried Fouvry[b], Véronique Aubin[a]**

a. Université Paris-Saclay, CentraleSupélec, ENS Paris-Saclay, CNRS, LMPS - Laboratoire de Mécanique Paris-Saclay, 91190, Gif-sur-Yvette, France.
b. Laboratoire de Tribologie et Dynamique des Systèmes, Ecole Centrale de Lyon, Ecully, France
c. Laboratoire Structures, Propriétés et Modélisation des Solides, CentraleSupélec, CNRS UMR 8580, Université Paris-Saclay, Gif-sur-Yvette 91190, France

vivien.lefranc@centralesupelec.fr, camille.gandiolle@centralesupelec.fr, siegfried.fouvry@ec-lyon.fr

**Keywords:** Fretting, Tribologically Transformed Structures (TTS), Wear, Ti-6Al-4V

## Abstract:

Fretting wear is a surface degradation process caused by oscillatory motion and contact slipping. During gross slip, high local stresses and plastic deformation in the surface and subsurface can lead to the creation of a nanosized grained structure called Tribologically Transformed Structure (TTS). The current paper studies the formation of TTS in an alpha-beta Ti-6Al-4V alloy under fretting loading while changing the contact pressure and the number of fretting cycles.

Cross-sections of wear scars are observed after polishing and chemical etching. Above a threshold pressure of 300 MPa, TTS appears early in the contact (before 1000 cycles) along with two other structures: a Third Body Layer (TBL) made of compacted debris and a General Deformed Layer (GDL) which is the plastic zone under the TTS. TTS first appears as islands and merges in the middle of the contact after enough cycles. Below 200 MPa, only TBL and GDL are formed. At 200 MPa, only small, localized TTS is found. All structures have the same chemical compositions as the initial bulk material except for the nitrided TBL. TTS has a very high hardness compared to the bulk. TTS was carefully extracted using a Focused Ion Beam (FIB) and its microstructure was observed with a Transmission Electron Microscope (TEM). It shows extreme grain refinement and is composed of two alternated zones. The first zone I is composed of α grains with a size of 20 to 50 nm with crystallographic texture. Zone II comprises nanosized equiaxed grains whose sizes range from 5 to 20 nm without texture. The results made it possible to establish a scenario of the appearance of the TTS according to the conditions of contact pressure and number of fretting cycles.

1. **Introduction**

Fretting is a surface degradation phenomenon that is caused by vibrational effects and slipping in contact [1,2]. It results in high maintenance costs and sometimes failure in industrial applications such as blade-disk contacts of airplanes' turbo-reactors [3,4]. Slipping can be separated into two regimes: partial slip where the contact center is stuck and surrounded by a sliding region [5], and gross slip in which the whole contact undergoes sliding. For partial and gross slip, the primary surface degradation mechanism is cracking and wear respectively [6]. Under gross slip and in a dry contact, fretting wear is associated with high local stresses and plastic deformation in the vicinity of the surface which can lead to the creation of a superficial layer called Tribologically Transformed Structure (TTS) [7,8]. It is formed in the early cycles of fretting and is destroyed, later giving birth to the third body which then controls the wear of the contact [9]. To assess the impact of TTS on wear, it is of prime importance to understand TTS characteristics and how it is formed.

TTS has been observed in multiple alloys (Titanium, Steel, Aluminium); it is characterized by its uniform white appearance after chemical etching and it is reported to be dozens of microns thick [7]. Mechanical characterization of TTS is complex: its low thickness limits mechanical testing; as such, micro-indentation and nano-indentation have been the main choices to characterize TTS [8,10,11]. Using Vickers micro-hardness measurements, Sauger et al. [8] have shown that TTS has a very high hardness, often more than two times the hardness of the bulk material it originates from. Its high hardness is thought to be due to strain-hardening mechanisms [10,11]. Often full of cracks, TTS is very brittle [12]. Furthermore, regardless of the initial phases, TTS is made of the most thermodynamically stable phase at ambient temperature which is austenite for steels and the α hexagonal closed packed phase for titanium alloys [7,8]. Moreover, TTS has the same chemical composition as the samples before fretting, as reported by Sauger et al. who performed energy dispersed X-ray (EDX) analyses on Ti-6Al-4V [13]. Concerning the TTS of Ti-6Al-4V, a previous study by Fayeulle et al. [7] highlighted a transformation of the subsurface after 100 000 cycles of fretting under ball-on-flat contact where a superficial TTS layer was formed of 20-50 nm grain diameter with no beta phase.

Due to the high number of parameters influencing fretting [14], the origin of TTS formation is still not understood. Temperature effects [15], material transfer [16] and high plastic strains [7,8,11] are all considered potential forming mechanisms. While the overall contact temperature seldom rises by more than 100°C in fretting tests [17], very localized flash temperatures at asperities of the contact could significantly increase the temperature followed by rapid quenching in the substrate [15,18,19]. Rigney [16] introduced the concept of Mechanically Mixed Layer (MML), a nanosized grain layer produced by mechanical mixing, that needs either reaction with the environment for homogeneous contacts or material transfer for heterogeneous contacts under fretting to form. As such, the chemical composition of MML is different from the bulk. The formation mechanism of MML is different from that of TTS which has the same chemical composition of the bulk and is not due to material transfer but comes from in-bulk transformation. Y. Liu et al. [20] studied the microstructure development of Ti-6Al-4V homogeneous contacts after sliding wear in vacuum and found that a 70 μm thick layer with a grain size of 50 to 100 nm is formed. It indicates that the reaction with the environment to produce nanometric grain layers akin to TTS. Sauger et al. [8] proposed that the accumulation of cyclic plastic deformation induced by fretting could create TTS. During fretting, a critical plastic deformation could be obtained in the surface with an associated dissipated energy. It could result in the creation of TTS through recrystallization processes [8,11]. More recently, L. Xin et al. [21] investigated TTS formation under fretting loadings using 304SS steel ball sliding against a flat specimen of 690TT alloy. They revealed that TTS is formed as a consequence of dynamic recrystallization as suggested by Sauger [8] and another nanograined structure is obtained from material transfer and mechanical alloying such as the MML introduced by Rigney [16]. As such, it seems possible for both mechanisms to happen in the same contact.

Microstructure characterization of TTS is complex, its nanosized grains cannot be observed by Scanning Electron Microscope (SEM) or Electron Backscatter Diffraction (EBSD) and needs the use of transmission electron microscope (TEM) to provide information on grains morphology, size and texture. Also, the fretting loadings influencing the formation of the TTS are not fully understood. As the contact pressure changes during fretting [22], TTS is not present in the whole contact zone and its localization should be investigated through cross section analysis and precise TEM sample extraction techniques. In this work, the formation of TTS in an alpha-beta Ti-6Al-4V alloy under fretting with flat-on-flat homogeneous contacts is studied while modifying the contact pressure and the number of fretting cycles. The cross-section subsurface is observed after polishing and chemical etching in order to view the evolution of the TTS layer. The microstructure of TTS at different states of transformation will be thoroughly examined using TEM images and diffraction patterns on samples extracted with FIB lift-out method.

## 2. Material and fretting experiment

### 2.1. Material

The material used in this study is Ti-6Al-4V alpha/beta alloy. It was maintained at a temperature of α–β domain, then quenched in water and annealed at a temperature of 700 °C. This alloy is commonly used in the aerospace industry thanks to its high specific strength and low density. The microstructure obtained is shown in figure 1, it is formed of 20 to 50 μm α nodules separated by α + β lamellae. The properties of this alloy are recorded in table 1.

### 2.2. Experimental method

#### 2.2.1. Test apparatus

The fretting test apparatus was mounted on an MTS hydraulic machine [23]. Tangential displacement was applied with a hydraulic actuator. The bottom sample was fixed below the hydraulic actuator axis. The angle and the position of the bottom sample can be adjusted before the test and the alignment between the surfaces was verified using a pressure paper. The normal force was applied before applying the tangential displacement and adjusted manually during the test if needed. The tangential force (Ft) and normal force (Fn) were recorded throughout the tests using force sensors and tangential displacement (δ) using a displacement sensor. A graphical representation of the testing apparatus can be seen in figure 2.

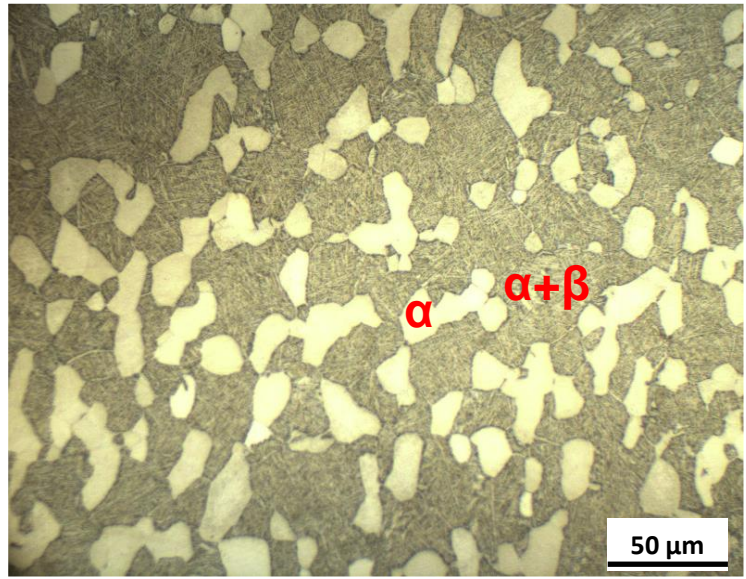

Fig. 1. Micrograph of the initial microstructure of the studied Ti-6Al-4V alpha/beta alloy: white grains are alpha phase grains surrounded by a lamellar structure of alpha and beta phases.

| Elastic Modulus E (GPa) | 119 |
|---|---|
| Poisson's coefficient ν | 0.29 |
| Yield Stress (MPa) | 970 |
| Density (g/cm$^3$) | 4.4 |

Table 1. Material properties of the studied Ti-6Al-4V alloy [24].

### 2.2.2. Test parameters

This study focuses on the effect of the contact pressure change and the number of cycles on the fretting wear response of flat-on-flat geometry. The tests were performed at ambient temperature (25 °C ± 5 °C) and relative humidity (RH=40 % ± 10 %). The frequency is fixed at 10 Hz and the sliding amplitude $\delta_0$ at 100 μm ($\delta_0$ is defined in figure 4) to maintain a gross slip condition in all the configurations. This results in a sliding speed of 4 mm/s. The bottom and top samples are 4 or 5 mm wide Ti-6Al-4V bars, the top sample is set perpendicular to the bottom sample. It results in a square contact area of 16 mm² (4x4 mm²) for the 4 mm bars and 25 mm² for the 5 mm ones with a surface roughness (Ra) < 0.4 μm. The applied normal force was varied from 1250 N to 5000 N for the 5 mm bars (resulting in contact pressures $p_0$ from 50 MPa to 200 MPa) and for the 4 mm bars the applied normal force ranged from 3200 N to 8000 N (respectively an initial contact pressure of 200 MPa to 500 MPa) for 20000 cycles. Further tests were done at $p_0$ = 300 MPa and $p_0$ = 400 MPa with a variable number of cycles from 1000 to 15000 to investigate fretting wear kinetics and its impact on TTS. The full scope of the studied parameters is illustrated in figure 3.

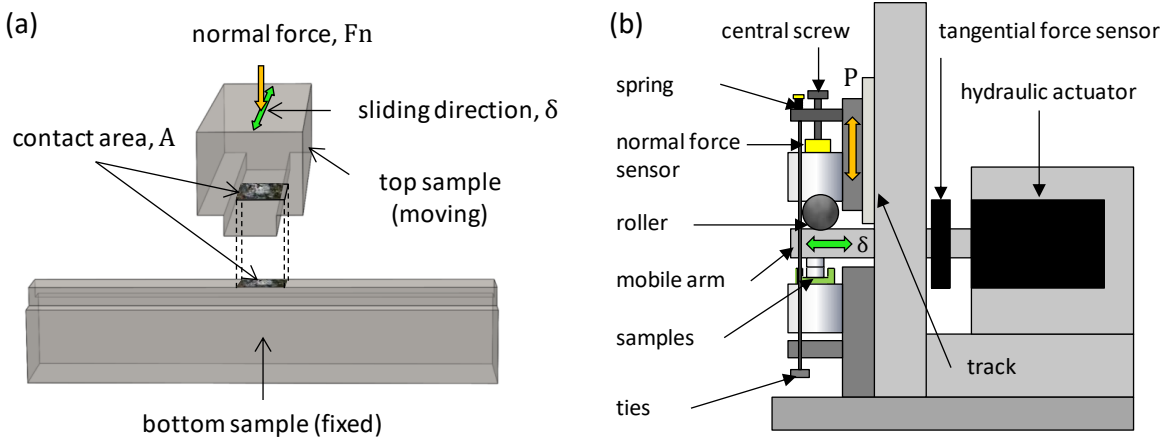

Fig. 2. Schematic presentation showing the (a) crossed homogeneous flat-on-flat contact and the (b) hydraulic fretting wear test system at Laboratoire de Tribologie et Dynamique des Systèmes (LTDS) [23].

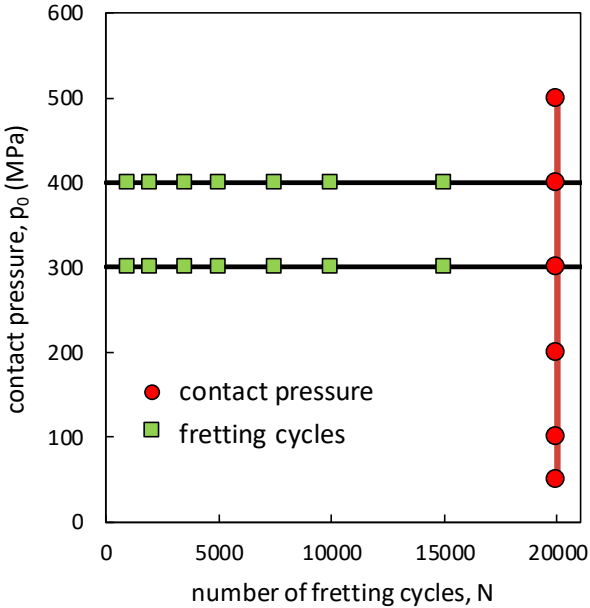

Fig. 3. Contact pressure values versus the number of cycles for the fretting tests. The frequency was fixed at 10 Hz and the sliding amplitude at 100 μm for each test.

2.3. Surface analysis: wear and EDX

After each test, the surfaces were cleaned in an ethanol ultrasonic bath for 20 minutes to remove all the superficial debris from the fretting scar. A white light interferometer VEECO Wyko NT 9300 was used in Vertical Shift Interference (VSI) mode with a x2.5 objective to get the topography of both surfaces. The wear volume was evaluated using the methodology described in [23]. A reference plane is identified corresponding to the height of the intact surface. The missing volume located below this reference plane is the volume of material removed from the fretting scar $V^-$ and the volume of material above the reference plane is the volume of material transfer on the interface $V^+$. The wear volume of one surface ($V^t$) is expressed by $V^t = V^- - V^+$. By summing both surfaces' wear volume, the total wear volume (V) is obtained.

To quantify wear damage, the energy wear rate (α) was introduced by Fouvry et al. [25] and derived from Archard wear law [26]. It highlights the linear relationship between the wear volume (V) and the cumulated dissipated friction energy $E_d$ (sum of the area of fretting cycles as in figure 4) in the contact. It is given by:

$$V = \alpha\left(\left(\sum E_d\right) - E_{dth}\right) \text{ when } \sum E_d > E_{dth} \text{ and } V = 0 \text{ if } \sum E_d < E_{dth} \qquad (1)$$

$E_{dth}$ is a threshold energy required to form TTS on the superficial layers of the surfaces and to initiate wear. Furthermore, EDX maps were acquired in a HELIOS 660 Dual-Beam SEM-FIB equipped with a EDAX Octane super 30mm² EDX detector and TEAM software. The surfaces were cleaned before analyses to characterize the compacted debris forming the Third Body Layer (TBL).

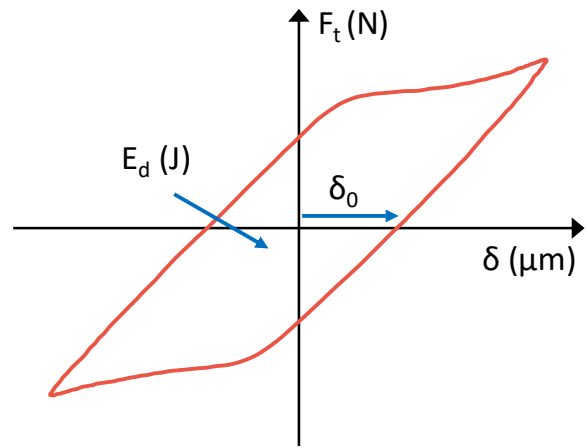

Fig.4. Reference fretting cycle of the tangential force ($F_t$) versus displacement amplitude ($\delta_0$).

2.4. Cross-section preparation and characterization

After the fretting tests, cross-sections of the wear scars were carried out. The specimen was cut in the direction of the fretting displacement near the middle of the scar using a diamond saw. The sample was then embedded with hot resin to ease polishing and to protect the fretted surface. Coarse to fine grinding was used on the cross-section using SiC abrasive papers followed by polishing with a 0.06 μm colloidal silica suspension to obtain a mirrored surface in the middle of the fretting scar. Subsequently, the cross-section was chemically etched using Kroll's reagent ($H_2O$ + $HNO_3$–HF) to reveal the microstructure. Optical images were done using ZEISS Axioscope 5 optical microscope and Scanning Electron Microscope (SEM) HELIOS 660 Dual-Beam SEM-FIB for both imaging and EDX point measurements at 15 kV.

Hardness analysis in the cross-sections was made using an NHT3 nanoindenter from Anton Paar mounted with a diamond Berkovitch tip. The Vickers hardness was determined using the Oliver and Pharr method [27]. Each indent was performed at a maximum load of 20 mN, a constant strain rate of 0.1 $s^{-1}$ and a holding time at maximum load before unloading of 10 seconds. Three valid measurements at each depth were taken to average the Vickers hardness at a given depth and are used to calculate the standard deviation. As the TTS can be full of cracks, indents made on cracks were not included in the hardness analysis.

To evaluate TTS microstructure, two Transmission Electron Microscope (TEM) samples were extracted and thinned with a Focused Ion Beam (FIB) using a Thermofisher HELIOS 660 Dual-Beam SEM-FIB with the in-situ lift-out method which is described in [28]. Both samples were observed in a TEM JEOL JEM-2100Plus with an acceleration voltage of 200 kV and equipped with a Gatan camera CCD CMOS RIO 16 4k. In addition to the high-resolution microstructural observations offered by TEM imaging, diffraction patterns were acquired to identify the phases and to check the presence of texture. Analysis of the diffraction patterns was done using CrysTBox software [29].

## 3. Results

### 3.1. Wear

The total wear volume V is plotted as a function of the number of cycles for each test in figure 5a. The tests at 20 000 cycles and the contact pressures of 100 MPa and 200 MPa acting as reference tests were duplicated revealing insignificant variation in the wear volume. Linear evolution is observed for both 300 MPa and 400 MPa with, as expected, higher wear volume for 400 MPa.

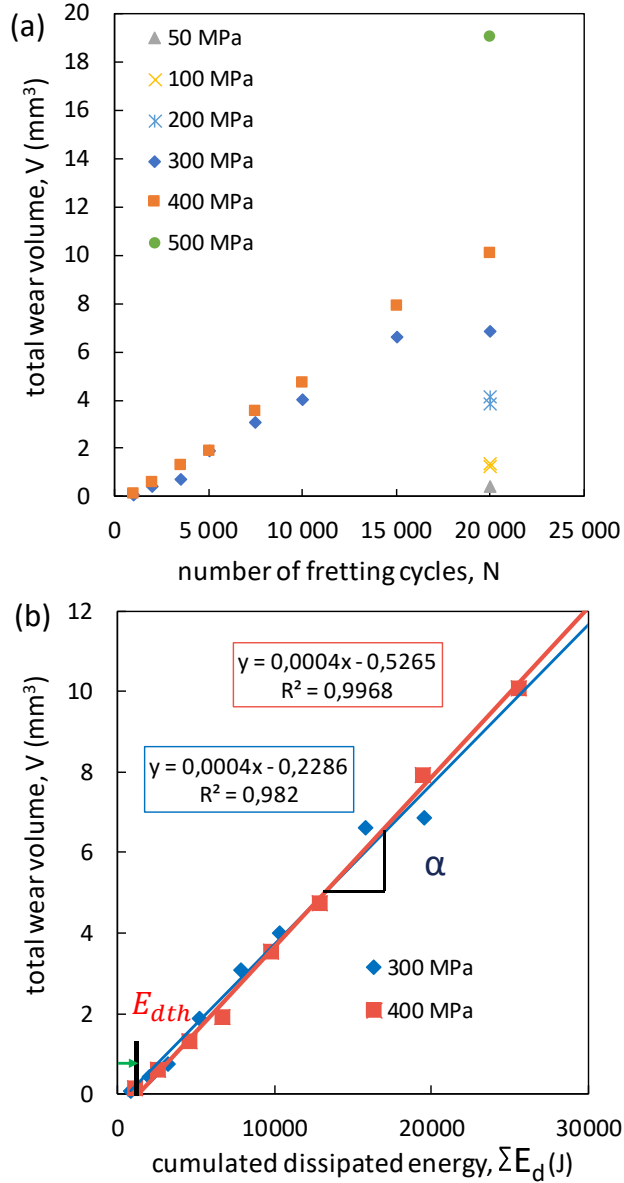

Fig.5. a) Total wear volume as a function of number of fretting cycles for all tests, b) total wear volume as a function of cumulated dissipated energy for 400 MPa tests.

The analysis of the different contact pressures after 20000 cycles confirms that the higher the contact pressure, the higher the wear volume. A better comparison can be achieved when comparing wear volume versus cumulated dissipated energy. Figure 5b confirms the linear increase of the wear volume versus the cumulated dissipated energy for both 300 MPa and 400 MPa satisfying equation (1) with α(300 MPa) = α(400 MPa) = $4.10^{-4}$ mm$^3$/J, $E_{dth}$(300 MPa)= 572 J. $E_{dth}$(400 MPa)=1316 J. For these conditions, the wear coefficient is the same as expected. Unfortunately, this energy wear analysis cannot be generalized for all contact pressures due to experimental costs. However, the energy wear rate can be extrapolated considering the tests at 20000 cycles: $\alpha_{(20k)} = \frac{V(20k)}{\sum E_d(20k)}$. Figure 6 plots the corresponding $\alpha_{(20k)}$ as a function of the contact pressure.

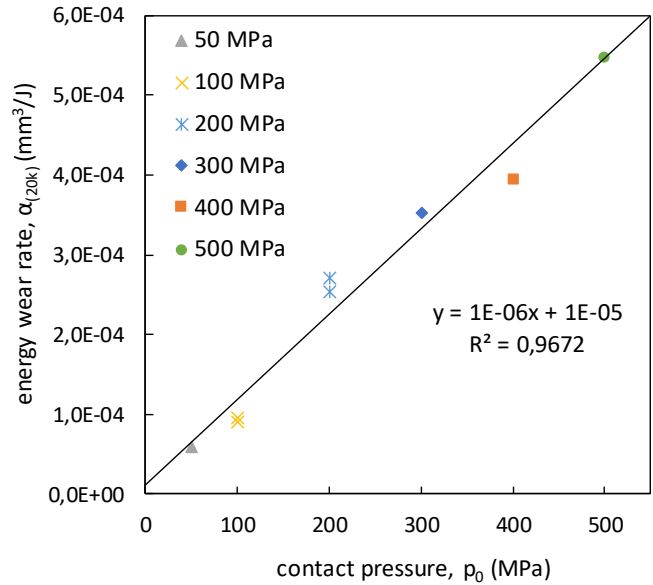

Fig.6. Energy wear rate versus contact pressure.

Surprisingly, no constant value was observed, but a linear tendency of the wear rate with the applied contact pressure. This fluctuation of the energy wear rate suggests a significant variation of the wear process as a function of the applied contact pressure. This aspect will be deepened in the following surface scar expertise.

3.2. Surface and debris analysis

Figure 7 compares the optical and EDX mapping of the fretting scars after surface cleaning for the 300 MPa contact pressure and fretting cycles from 1000 to 7500. The EDX maps of the fretting scars are shown with the nitrogen, oxygen, and titanium content in blue, green, and red respectively. When focusing on the medium test duration of 2000 cycles (figure 7b), it is interesting to observe a composite structure inside the fretting scar. The inner part circled with a red dash-line, is poor in oxygen and nitrogen, implying direct metal-metal interaction. This central zone is surrounded by a yellow-like area characterized by higher nitrogen content. Such structure corresponds to the activation of the nitriding process as detailed by C. Mary et al. [30]. Finally, the outer part is composed of a high concentration of black oxide debris. This concentric structure at the fretted interface is consistent with the contact oxygenation concept modelled [31] and successively simulated using an advection diffusion reaction model introduced by Baydoun et al. [32]. One interesting aspect is the dynamic evolution of such structure. At the beginning of the test, the metal-metal domain is preponderant, with the oxidation and nitriding process restricted in a narrow band at the outer parts of the contact (figures 7a and 7d). Then, the inner metal-metal domain decreases in size whereas the nitriding domain spreads towards the central part of the contact (figures 7b and 7e). Finally, after N=7500 cycles, the inner metal-metal domain entirely disappears and the whole interface is covered by nitride debris layer with a narrow oxidation domain in the outer part of the domain (figures 7c and 7f).

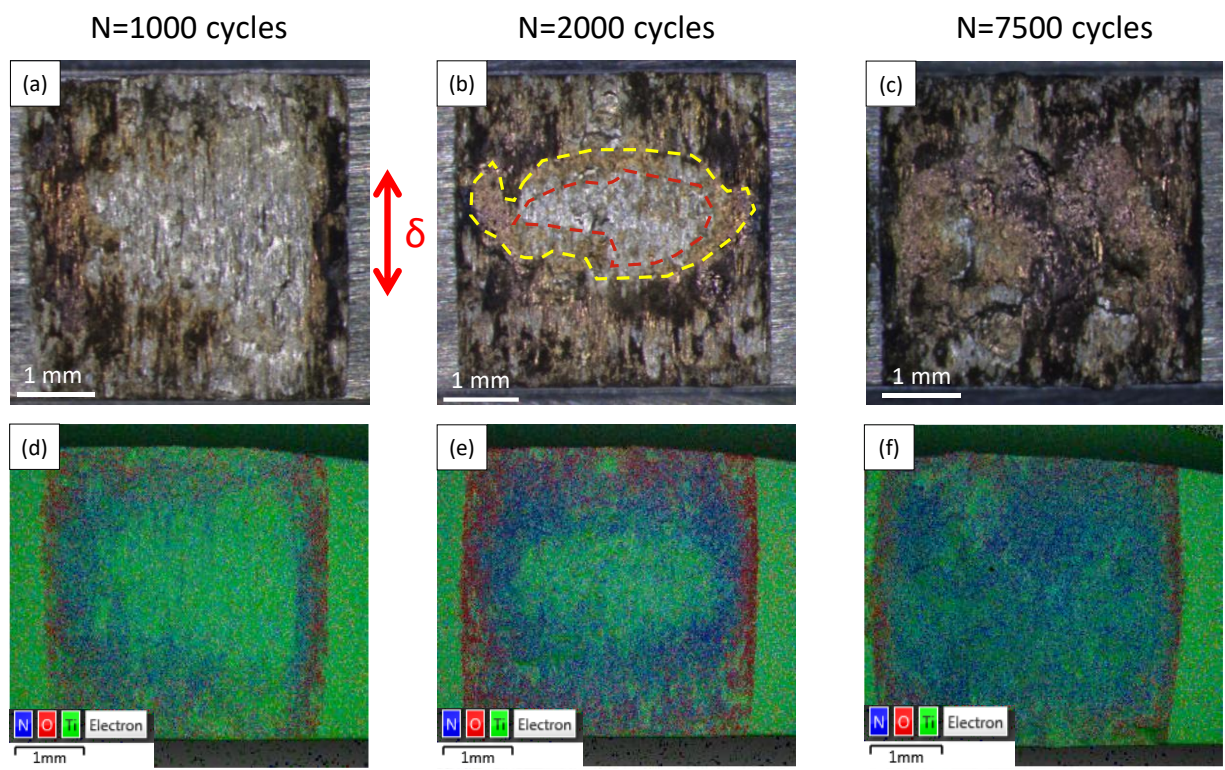

Fig.7. Optical pictures the bottom specimen after fretting tests under 300 MPa and number of cycles N of a) 1000, b) 2000, c) 7500 and their related EDX maps d), e), and f) respectively.

To better interpret these results, cross-section observations combined with EDX mapping and nano-indentation have been performed.

### 3.3. Cross-section analysis

Figure 8 illustrates the cross-section after chemical etching of the fretting scar cross-section obtained under 300 MPa contact pressure and 2000 fretting cycles. It was taken in the center of the contact parallel to the sliding direction. In the inner part of the contact a uniform white layer, up to 60 µm in thickness, is seen: the Tribologically Transformed Structure (TTS) (figure 8b). Below this layer, a plastically deformed zone is detected referred to as the General Deformed Layer (GDL) [33] which has a thickness of 20 to 30 µm. Underneath the GDL, the undeformed bulk is found. On the outer part of the surface, the Third Body Layer (TBL) is observed and is composed of agglomerated power debris (figure 8c and 8d). The TBL is formed with a thickness of approximately 5 to 15 µm.

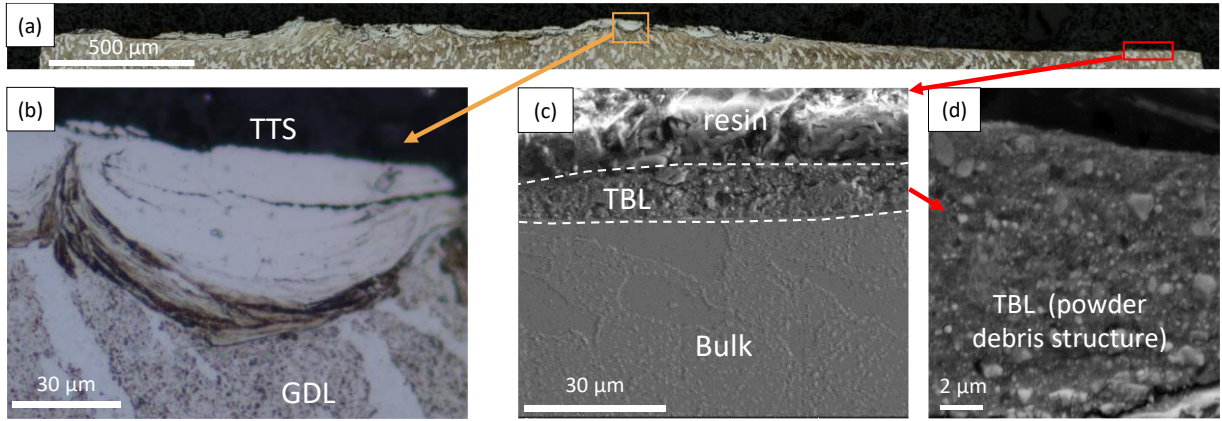

Fig.8: Cross section of the fretting test with a contact pressure of 300 MPa and 2000 fretting cycles. Optical micrographs (a) of the overall contact, (b) with a zoom on TTS. (c) Secondary electron image of the outer contact zone with (d) a zoom on TBL.

Semi-quantitative EDX point analyses were performed on TTS, TBL and GDL. The content of nitrogen, aluminium, titanium, and vanadium in TBL, TTS, GDL obtained with this analysis are shown in weight percent in table 2.

| Element | TBL | TTS/GDL | Bulk (supplier data) |
|---|---|---|---|
| N (Wt %) | 9.72 | 0 | <0.05 |
| Al (Wt %) | 5.77 | 5.92 | 5.5-6.75 |
| Ti (Wt %) | 82 | 89.5 | Base |
| V (Wt %) | 2.51 | 4.58 | 3.5-4.5 |

Table 2: Content of nitrogen, aluminium, titanium, and vanadium in TBL, TTS, GDL and bulk (300 MPa contact pressure and 2000 fretting cycles).

Table 2 reveals a higher nitrogen content in the TBL layer, which indicates that nitriding of titanium occurs. As nitrogen is difficult to quantify with precision using EDX, these results should be used as a qualitative evaluation of the presence of nitrogen in the TBL layer. The TBL layer can be linked to the yellow nitride zone seen on the scar surface (figure 7). Cross section observation of the outer part of the contact, where higher oxygen content exists, could not allow the observation of the TBL. The oxidized debris particles appear more powdery and less cohesive than the nitride debris and therefore are more easily ejected from the interface. As for TTS, the EDX measurements show no difference in composition from the bulk GDL or bulk which would indicate that this TTS was not formed due to reaction with the atmosphere.

Nano-indentation hardness assessment was performed to characterize the mechanical behaviour of the TTS and GDL. The resulting hardness measurements are shown in figure 9. Very high Vickers hardness is found in the TTS with $H_{TTS} = 1170$ HV which is in accordance with previous literature results [7,8]. In comparison, the hardness in the bulk is $H_{bulk} = 560$ HV which is typical of the Ti-6Al-4V alloy. Interestingly, the hardness of the GDL layer is lower than for the bulk with $H_{GDL}$ = 420 HV. This lower hardness value could be explained by the plastic softening largely described in the literature for Ti-6Al-4V under cyclic loadings [34] and fretting [35].

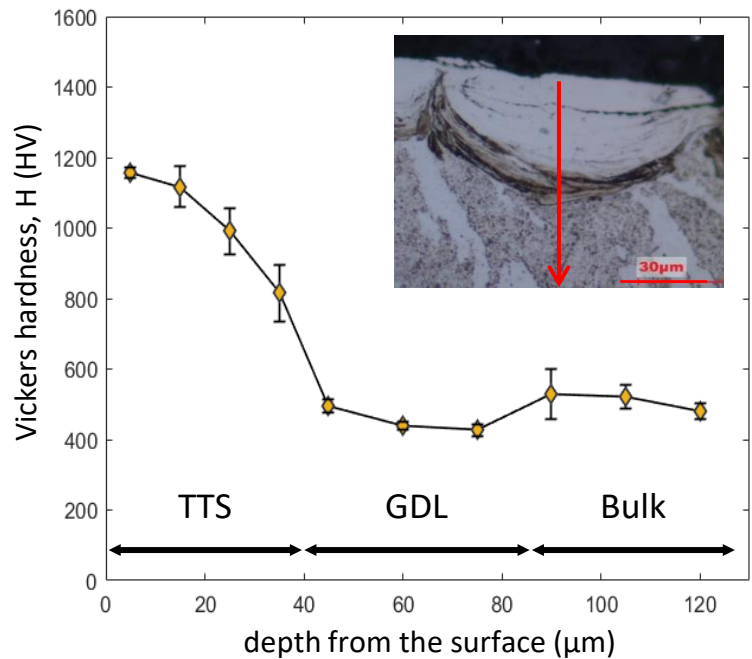

Fig.9: Vickers hardness as a function of the depth from the surface (300 MPa contact pressure and 2000 fretting cycles).

These observations were done on the reference case studied for a contact pressure of 300 MPa after 2000 fretting cycles. The stabilized response of the interface after 20000 cycles shows that the Third Body Layer (TBL) and the General Deformed Layer (GDL) are present in all contacts regardless of the contact pressure (figure 10). However, Tribologically Transformed Structure (TTS) activation is observed only for contact pressures larger than 200 MPa. At a contact pressure of 200 MPa, TTS is observed in the form of small, isolated islands or needles (figure 10a), which can be observed under the TBL layer. Above 300 MPa, massive TTS are observed mainly in the center of the contact. It should be noted that neither lateral extension nor thickening of the TTS layer are necessarily induced by an increase in the contact pressure.

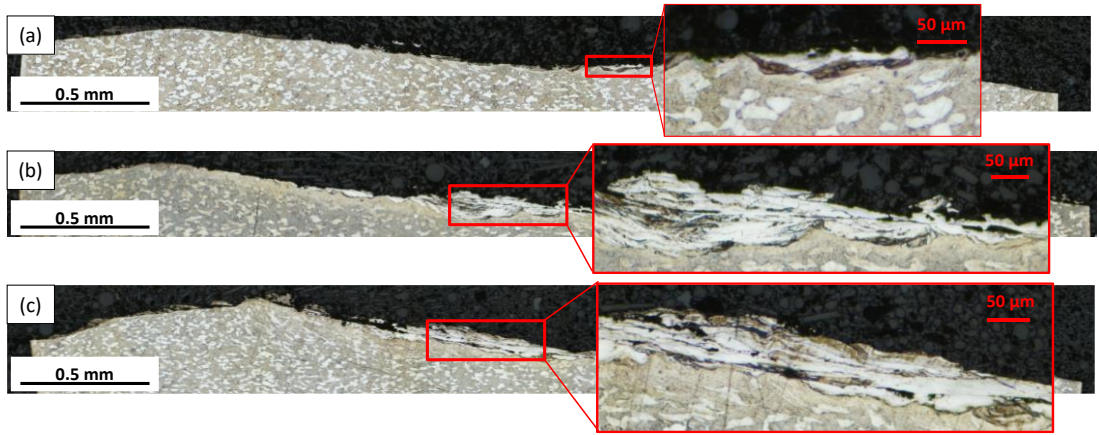

Fig.10: Optical images of the fretting scar cross-section after 20000 cycles for a contact pressure of a) 200 MPa, b) 300 MPa and c) 400 MPa.

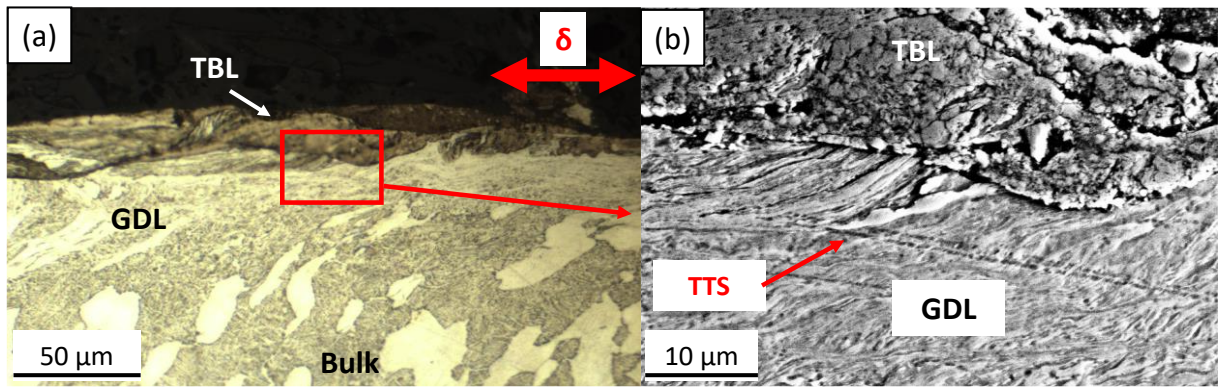

Fig.11: a) Optical and b) BSE images of cross-section observation with highlighted lamellar TTS (200 MPa contact pressure and 20000 fretting cycles).

A dynamic analysis of TTS formation for a representative contact pressure of 300 MPa at a number of fretting cycles ranging from 1000 to 20000 shows a continuous process of TTS zone extension. Interestingly, TTS often appears through disconnected islands at low number of cycles (figure 12) expanding then to the center of the contact as a single body (figure 10b). This expansion happens before the TBL forms on the entire contact (after 7500 cycles as shown in section 3.2), which confirms the EDX analysis outlining that the TTS transformation occurs in the first bodies. It should be noted that systematic EDX and indentation analyses were performed on all the TTS identified by chemical etching (white traces) with similar results.

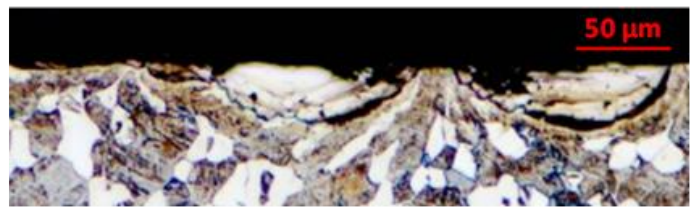

Fig. 12: Optical image of TTS island on the cross-section of a contact tested at a contact pressure of 300 MPa at 1000 fretting cycles.

### 3.4. Microstructure analysis of TTS

The TTS found in the form of needles formed under a contact pressure of 200 MPa and the massive TTS formed at 300 MPa were investigated with a Transmission Electron Microscope (TEM) JEM-2100Plus with an acceleration voltage of 200 kV, equipped with a Gatan camera CCD CMOS RIO 16 4k. Two samples were extracted and thinned with a focused ion beam (FIB) using a HELIOS 660 Dual-Beam SEM-FIB. First, sample A was taken in the TTS formed under 200 MPa and 20000 cycles with a localized lamellar aspect and surrounded by GDL (figure 13a). Then, sample B was extracted from a massive TTS obtained under 300 MPa during 1000 cycles (figure 13a).

The TEM lamella A involved both GDL and TTS domains (figure 13b). TEM image of GDL zone displays grain sizes from 70 to 150 nm with elongated shapes (figure 13c). By contrast, the TTS microstructure is formed of grains ranging from 20 - 50 nm (figure 13e). The diffracted areas are materialized in both images with a red circle. The diffracted patterns associated with the two zones observed are shown in figures 12d and f. In each case, only α planes were diffracted with no presence of β phase in the diffraction patterns indicating that a phase transformation from α+β to α occurred during the fretting test. Another noteworthy aspect of the diffraction patterns is the presence of arcs of circle in both zones related to texturation. If the number of grains diffracted is statistically representative, the diffraction pattern of a randomly oriented polycrystalline structure should only contain full, homogeneous circles. In case of a textured polycrystal, in which grains have a preferred orientation, arced diffraction patterns occur [36,37]. In the GDL zone, the grain size is relatively large compared to the diffracted zone as seen in figure 13c, as such, it is possible that the number of grains in the diffracted area of this zone is too small to be a representative volume element. It is thus not possible to estimate if this zone is textured or not because of the fragmented aspect of the diffraction pattern (figure 13d). Nonetheless, the grain size in the TTS domain is smaller, allowing enough grains to be diffracted in the diffraction pattern (figure 13e), as seen by the smoother arced pattern seen in figure 13f. Considering that the arced diffraction patterns are being more defined, it can be concluded that there is a crystallographic texture in this TTS zone.

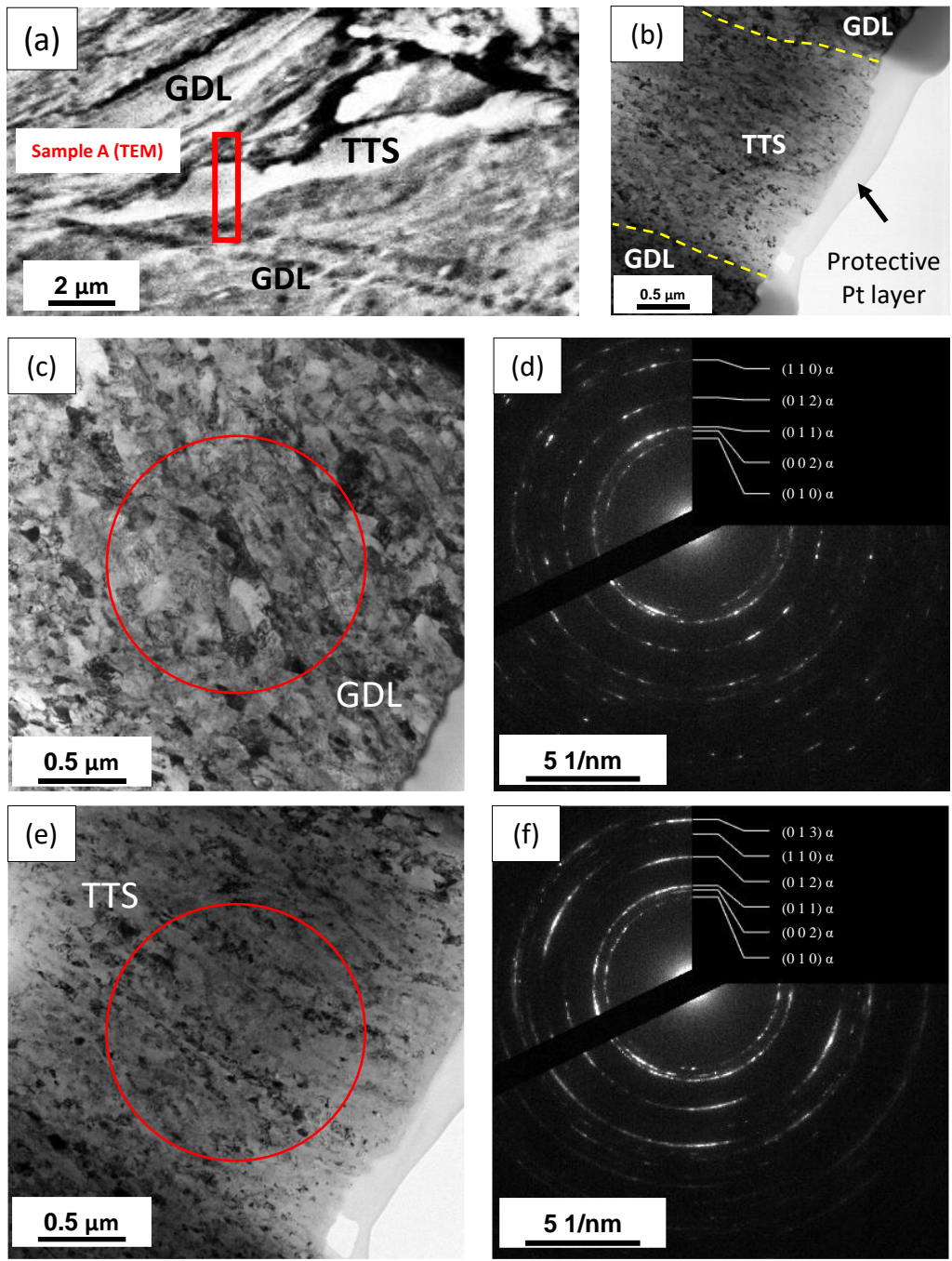

Fig. 13: a) BSE image of the extraction zone of sample A. TEM bright field images of b) the overall view of sample A, c) the GDL and e) the TTS with d) and f) being the associated diffraction patterns. Diffracted areas are circled in red.

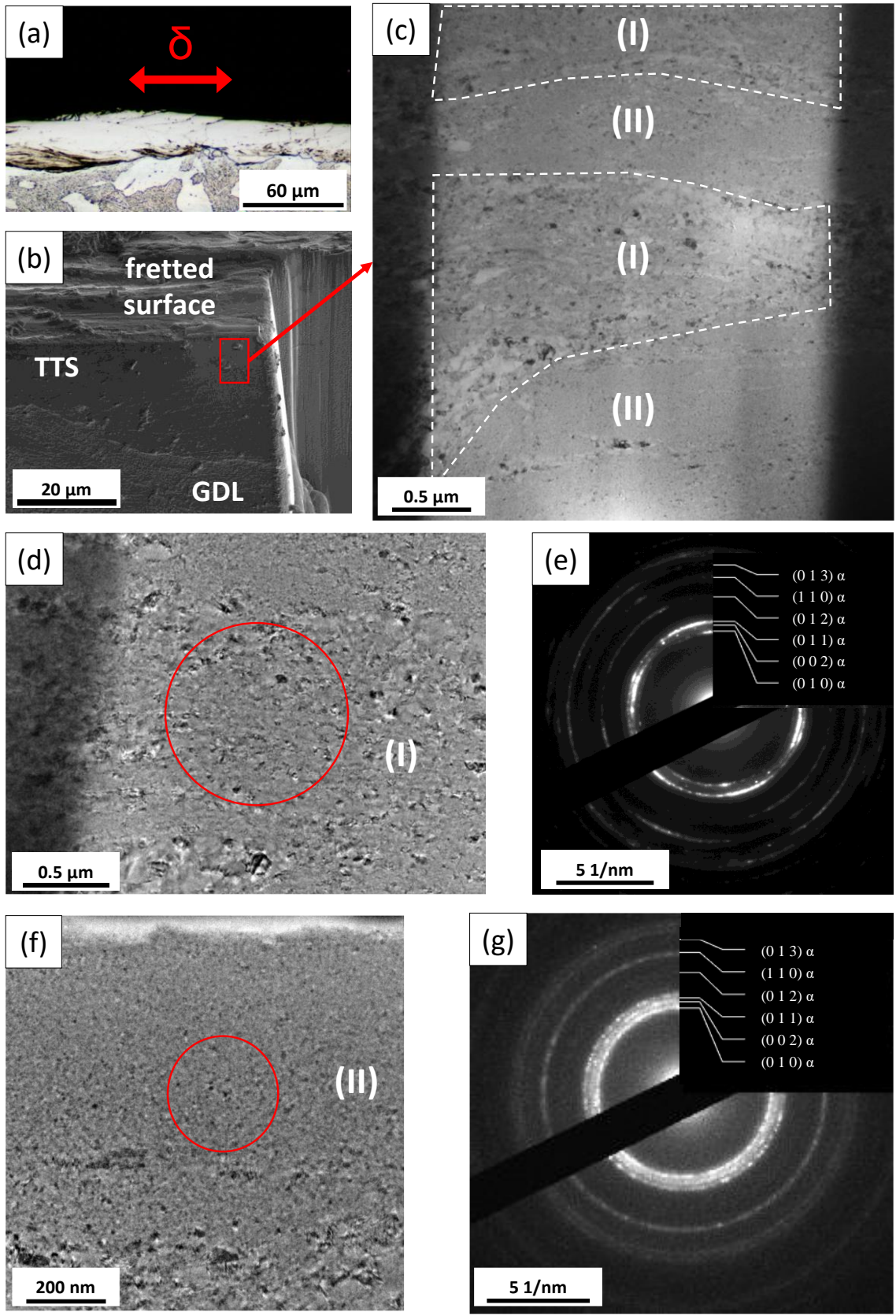

Fig. 14: a) Optical and b) BSE images of the specimen tested under 300 MPa during 1000 fretting cycles with position of the TEM sample extraction (sample B). c) Bright field image of TEM lamella B. d) TTS (I) f) TTS (II) with e) and g) the associated diffraction patterns. Diffracted areas are circled in red.

The specimen tested under 300 MPa during 1000 fretting cycles shows a well-defined TTS zone (figure 14a). Sample B was extracted from it (figure 14b). TTS is in fact not homogeneous but displays a "sandwich" structure made of alternated "coarse" TTS structure (I) and an ultrafine TTS domain (II) which appears uniformly grey in figure 14c. Grain zone (I) is composed of grains from 20 to 50 nm (figure 14d) with only alpha phase as seen in the figure 14e. The diffraction pattern shows partial arc circles suggesting crystallographic texture. It seems that coarse TTS (I) observed in the massive TTS structure (sample B) is similar in many aspects to the localised TTS described in specimen A. They both present a similar grain size distribution from 20 to 50 nm and are made of only alpha phase with clear texturation. The ultrafine domain (II) is composed of equiaxed nanosized grains of size ranging from 5 to 20 nm as seen in figure 14f. The diffraction pattern associated with this zone (figure 14g) shows that only alpha structure is present. The diffraction circles in this zone are uniform and continuous, TTS (II) appears to have an isotropic crystallographic texture.

To conclude, the ultrafine TTS (II) formation corresponds to a more advanced state of transformation of TTS and is generated only at very severe plastic loading conditions. It is consistent with the fact that TTS (II) was only observed in massive TTS, which was present for higher contact pressures (>200 MPa) in the current test conditions. The originality work of this shows that TTS is not homogeneous but in fact heterogeneous with an alternation of two types of layers: (I) a less than 50 nm grains textured layer and (II) an isotropic very fine (less than 20 nm) grain layer.

## 4. Discussion

This study comprised fretting tests of flat-on-flat homogeneous Ti-6Al-4V contacts. Surface analysis shows that when increasing the number of fretting cycles, the nitrided Third Body Layer (TBL) spreads from the borders of the contact to the inner zone. Similar nitriding processes were already observed during fretting in titanium alloys [24,30,38]. It can be explained by the contact oxygenation approach developed by S. Baydoun et al. [31].The oxygen partial pressure drops during fretting tests from the borders of the contact to the center due to fretting-induced oxidation processes on the borders. Below a particular oxygen partial pressure threshold, the oxygen does not react with the titanium, instead, the partial pressure of nitrogen rises, and titanium can react with it. This oxygen partial pressure threshold is also linked to the abrasive and adhesive wear domains. The external contact zone, where the oxygen partial pressure is above the threshold value, undergoes abrasive-oxidational wear. On the other hand, adhesive wear prevails in the internal contact zone where the oxygen partial pressure is below the threshold value. Using a wear model integrating the third body approach with the contact oxygenation concept [39], Arnaud and co-workers marked an increased contact pressure in the inner part of the contact where adhesive wear is activated. This could explain why the TTS induced by plastic deformation is mainly observed in the central part of the contact.

Although nitriding and oxidation processes happen during fretting, TTS does not seem linked to environmental interaction as its chemical composition is the same as the bulk material. Thus, it seems unlikely that the TTS is formed of compacted debris or mechanical mixing as in [16],

MML is not observed in the current study. The observed TTS formation directly happens in the first bodies and is coherent with the model proposed by Sauger et al. [8,13].

The dynamics of TTS formation and the coupled effect of the contact pressure were also studied. Figure 15 summarizes the evolution of these phenomena.

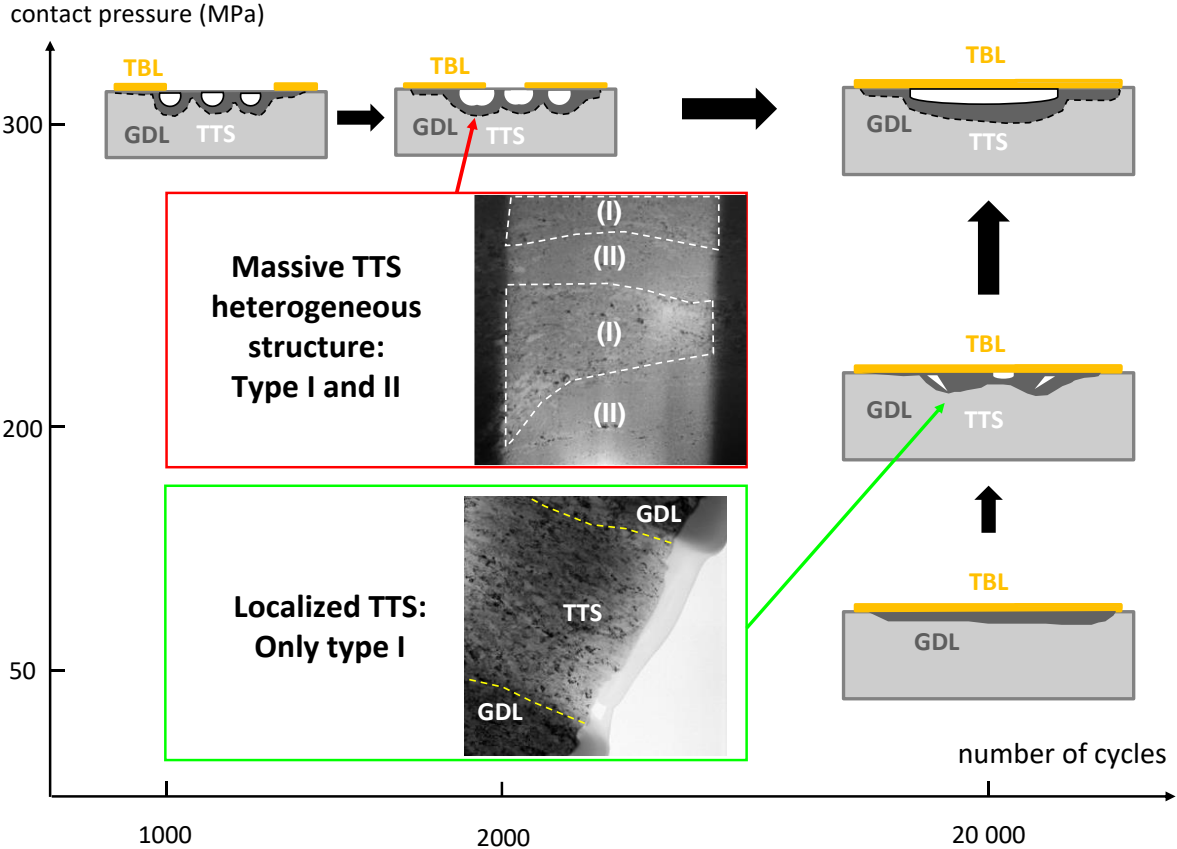

Fig.15: Coupled effect of the contact pressure and number of fretting cycles on the TTS, TBL and GDL formation.

Considering that a stabilized state is reached at 20000 cycles, it is shown that only the general deformed layer (GDL) and TBL are formed below a certain threshold contact pressure. The first phases of the TTS transformation are activated in very punctual layers above a certain loading threshold associated with $p_0 = 200$ MPa. The formation of a massive TTS is observed only from a certain plastic loading threshold associated with this study with an average contact pressure of 300 MPa. The formation of a massive TTS occurs in stages, with the nucleation of the first TTS seeds formed before 1000 cycles, which will tend to coalesce to form a macroscopically uniform TTS structure. This evolution can also be completed by a microstructural analysis of the latter. Indeed, the transformation of TTS is done in several steps. In the first step, the bulk, constituted of alpha-beta phase with a grain size of about 40 μm, undergoes a first transformation leading to a GDL structure in which all the beta phase is transformed into alpha phase and with a significant grain size reduction to reach 70 to 150 nm. This structure cannot be considered as TTS because it does not appear uniformly white after etching. Its hardness also reveals softening mechanisms very different from TTS. Above a loading threshold associated with a threshold pressure of 200 MPa, a second transformation process takes place to form a type I TTS (TTS zone within the GDL, therefore local transformation) made up solely of alpha phase with textured grain sizes of 20 to 50 nm. Finally, by increasing the number of cycles or the loading levels, an ultimate transformation of the material is activated. It involves the creation of a heterogeneous material with both the previous TTS zone (I) and an ultrafine isotropic structure with a grain size of 5-20 nm, called zone (II). Hence, between 200 and 300 MPa, there is only type (I) structure, whereas above 300 MPa, massive TTS is formed with alternating type I and type II structures. One can ask the question of the stability of this heterogeneous TTS and could eventually consider a complete transformation of the TTS into type (II), but it has not been observed in this study. A much larger number of cycles may be needed to reach this homogeneous state; however, the TTS is continuously destroyed and rebuilt during the fretting process so it might not be achievable. Additionally, it must be emphasized that the extreme grain shrinkage seen in TTS explains the high hardness of the TTS thanks to the Hall-Petch effect [40–42].

This work is based only on mechanical effects, but other effects may occur in the formation of TTS, for example, thermal effects. At the macroscopic level, and the frequency being low, the macroscopic temperature will not exceed 100°C. However, it should be noted that flash temperatures are confined to the scale of asperities of a few microns [18,19,43], but the layers of TTS formed are often greater than 50 μm and therefore are larger by a factor of 10 compared to the size of an asperity. An explanation based solely on very localized flash temperature makes it difficult to explain such TTS thicknesses. Thus, the hypothesis of plastic-induced transformations on the surface combined with high hydrostatic pressures, which prevent the fracturing of these very fragile structures, is more relevant. This hypothesis is reinforced by the observations made in severe plastic deformation (SPD). Indeed, the severe grain refinement found in this work is similar to those obtained in SPD processes. Ti-6Al-4V tested under high-pressure torsion [44] allowed structures with grain size below 100nm to form, although the resulting structure happens to be homogeneous. Also, Ti-6Al-4V under SPD processes underwent phase transformation from α+β towards α. As such, the beta phase disappearance in the GDL and in the TTS is akin to the dissolution of β phase reported for SPD processes on titanium alloys. Thus, plastic-induced transformation could explain the extreme grain size refinement and the change of phases inside the TTS even at room temperature.

In addition, these results are consistent with the wear kinetics that increase with pressure. Indeed, high pressures allow the formation of TTS. These very fragile structures facilitate the formation of debris particles which are powdery in nature and hence are easily evacuated from the interface. Debris from TTS are assumed to require low energy to be generated and to be removed from the interface. Moreover, for low pressures where only the GDL can be formed, the creation of debris is less energetically favourable and the debris particles, being more cohesive because of their metallic nature, require more energy to be ejected. The increase in the energy wear coefficient as a function of the contact pressure seen in figure 6 can be explained by the microstructural evolution of the interface. Further investigation to identify the two-threshold deformation state is needed for future accurate modelling of TTS formation.

# 5. Conclusion

In this work, the effect of the contact pressure and the number of fretting cycles on wear and TTS formation was investigated for flat-on-flat geometry on Ti-6Al-4V homogeneous contacts. When the contact pressure increases, different microstructure transformation stages are observed. After fretting, the first transformation step allows to obtain GDL with grain sizes ranging from 70 to 150 nm. At this point, the beta phase has already disappeared from the GDL and is not detected. The second step of the transformation is formed when localized TTS surrounded by GDL is observed at a contact pressure of 200 MPa. This TTS shows 20 to 50 nm grains and crystallographic texturation (type (I)). In the final transformation step, above $p_0 =$ 300 MPa, a sandwich-like structure is revealed in the massive TTS. Two separate zones are highlighted: one with grain size ranging from 20 to 50 nm with texturation type (I) (that is also seen in the localized TTS surrounded by GDL) and an equiaxed nanometric randomly oriented grained zone with grains ranging from 5 to 20 nm (type (II)). As such, TTS is a heterogeneous structure with extremely fine grains. These results made it possible to establish a scenario of the TTS appearance according to the contact pressure conditions and the number of fretting cycles.

# Acknowledgements

FIB-SEM work was carried out using the facilities available at the LMPS laboratory within the MATMECA consortium, which is supported by the ANR under the contract number ANR-10-EQPX-37.

# Declaration of Competing Interest

The authors declare that they have no known competing financial interests or personal relationships that could have appeared to influence the work reported in this paper.